\documentclass[%
 reprint,
 amsmath,amssymb,
 aps,prd
]{revtex4-1}

\usepackage{graphicx}
\usepackage{dcolumn}
\usepackage{bm}
\usepackage{tabularx}
\usepackage{amssymb}
\usepackage{amsthm}
\usepackage{amsmath}
\usepackage{caption}
\usepackage{setspace}
\usepackage{enumitem}
\usepackage{textcomp}
\usepackage[section]{placeins}
\usepackage[utf8]{inputenc}
\usepackage{wrapfig}
\usepackage{epstopdf}
\usepackage[toc,page]{appendix}
\usepackage{enumitem}
\usepackage{xcolor}

\newcommand{\beq}{\begin{equation}}
\newcommand{\eeq}{\end{equation}}

\begin{document}
\preprint{APS/123-QED}

\title{Asymptotic safety and Conformal Standard Model}

\author{Frederic Grabowski${}^1$}
\author{Jan H. Kwapisz${}^2$}%
\email{Corresponding author: Jan.Kwapisz@fuw.edu.pl}
\author{Krzysztof A. Meissner${}^2$}
 \affiliation{${}^1$ Faculty of Mathematics, Informatics and Mechanics, University of Warsaw ul. Banacha 2, 02-093 Warsaw, Poland}
\affiliation{${}^2$ Faculty of Physics, University of Warsaw
ul. Pasteura 5, 02-093 Warsaw, Poland}

\date{\today}

\begin{abstract}
We show that the Conformal Standard Model supplemented with asymptotically safe gravity can be valid up to arbitrarily high energies and give a complete description of particle physics phenomena. We restrict the mass of the second scalar particle to $\sim 300$ GeV and the masses of heavy neutrinos to $\sim 340$ GeV. These predictions can be explicitly tested in the nearby future.
\end{abstract}

\maketitle
\section{Introduction}
\noindent
In the recent years there were many extensions of the Standard Model (SM) proposed to deal with the SM drawbacks like triviality, too weak CP violation for baryogenesis, hierarchy problem and also with lack of dark matter candidates.
Examples of these extensions include Grand Unification Theories (GUT) \cite{PhysRevLett.32.438,BURAS197866}, supersymmetric models \cite{PhysRevD.24.1681,IBANEZ1981439,DIMOPOULOS1981150}, Higgs portal models \cite{Higgsportal,Shaposhnikov:2006xi} or the Conformal Standard Model \cite{MEISSNER2007,Latosinski2015,Lewandowski2017}.
 Usually free parameters of these models, like masses of  new proposed particles, are known neither theoretically nor experimentally. Possibility of narrowing them down, using some theoretical reasoning, to a small interval would be a huge advantage in the search for new particles and interactions. 
 
One way to restrict the low energy values of couplings is to include the gravitational corrections to the $\beta$ functions and demand the asymptotic safety condition (AS) on the running of renormalisation group equations (RGE) for given initial conditions \cite{ShaposWetterlich}. These conditions impose bounds on the model free parameters values, which are initial conditions for RGE. Then the model which satisfies them is UV fundamental. However this reasoning is valid only if gravity indeed has a non-perturbative asymptotically safe fixed point. This hypothesis \cite{Weinberg} is currently under investigation, however there are strong indications that it is really so, see for example \cite{Eichhorn:2017egq,Lauscher:2005xz,Salvio:2017qkx}. Much of the work is done using the non-perturbative Functional Renormalisation Group methods \cite{Wetterich:1992yh,Morris:1993qb,Morris:1994ie}. While the Wetterich equation for flowing action $\Gamma_k$ is exact (from which one gets the effective action $\Gamma= \lim_{k \to 0} \Gamma_k$), however it is very difficult to deal with. To solve it and hence find the effective action, one has to restrict himself to the finite array of couplings. This, non-perturbative, approach gave the promising results on mass difference of charged quarks \cite{Eichornquarks} or on explanation of the top mass \cite{Eichorntop}. On the other hand the perturbative approach was used to predict the Higgs mass \cite{ShaposWetterlich} with astonishing accuracy. 
\\
In this article we analyse the Conformal Standard Model (CSM) \cite{MEISSNER2007,Latosinski2015,Lewandowski2017}, which is of Higgs portal type, but has additional structure. This model extends SM by adding one new complex scalar field with its phase as a dark matter candidate and right-chiral neutrinos. The fundamental assumption which underlies this model and other similar models is that there is no new physics between the weak scale and the Planck scale. This mean that for example the masses of heavy neutrinos or vacuum expectation value of new scalars should be of order of 1 TeV. Indeed the observational abbreviations from the Standard Model such as neutrino masses and oscillations, dark matter and dark energy, baryon asymmetry of the Universe and inflation can be understood without introducing an ntermediate new scale, see for example \cite{Shapo1,Shapo2}.  The introduction of the gravitational contributions to the matter beta functions makes all the matter couplings go to non-interacting fixed point at roughly Planck scale, hence at least theoretically it is unnecessary to introduce the new degrees of freedom at some intermediate scale, see \cite{Tavares:2013dga}, in order to make Standard Model a CFT at high energies. Moreover the Large Hadron Collider (LHC) hasn't detected any discrepancies from the Standard Model, with no signs of supersymmetry. This is why the Higgs portal models and Conformal Standard Model attract a lot of attention as they can deal, in principle, with the drawbacks of the Standard Model without changing its structure deeply and adding only a few particles and interactions to the SM. Such models don't posses any higher dimension operators in the Lagrangian, since they are the negative dimensional operators. In the Functional Renormalisation Group they should be taken into account. Moreover they can can affect the Higgs sector however none of these effects have been confirmed yet, see for example \cite{Rosiek,Borchardt,Sondenheimer}). Hence in our article we deal (we truncate only to the renormalisable operators) only with those matter operators which are in the Lagrangian of the Conformal Standard Model.
\\
 By taking into account the gravitational corrections to the beta functions and using the AS conditions (and assuming that Weinberg hypothesis holds) we are able to calculate the allowed range of Higgs and the second scalar coupling parameters such that the CSM can be a UV complete theory. By taking into account the experimental LHC data and the model restrictions for the values of free parameters we are able to predict the allowed second scalar mass. We can also narrow down masses of right-chiral neutrinos.  One should also mention that there are some studies on models with asymptotically safe behaviour without taking into account gravitational corrections \cite{PhysRevLett.120.211803,Pelaggi:2017abg,Mann:2017wzh,Barducci:2018ysr}. 
\section{Model and method}
\subsection{Higgs portal models and Conformal Standard Model}
In this paragraph we briefly introduce the Conformal Standard Model. 
The CSM lagrangian \cite{Lewandowski2017} is given by
\beq
\mathcal{L} = \mathcal{L}_{kin} + \mathcal{L}_Y - V,
\eeq
with the following kinetic terms
\begin{eqnarray}
\label{the Conformal Standard Modellagrscalar}
\mathcal{L}_{kin} =  \mathcal{L}_{kin}^{SM} + (D_{\mu}H)^{\dagger}(D^{\mu}H) + (\partial_{\mu}\phi^{\ast}\partial^{\mu}\phi)  \nonumber \\ 
+i \overline{N}^j_{\dot{\alpha}} \overline{\sigma}^{\mu \dot{\alpha} \beta} \partial_{\mu}  N_{\beta}^j&,
\end{eqnarray}
where $H$ is the SM scalar $SU(2)$ doublet. The $\phi(x)$ is a gauge sterile complex scalar field carrying the lepton number, and couples only to gauge singlet neutrinos $N_{\alpha}^i$. The $\mathcal{L}_{kin}^{SM}$ contains all the non-scalar Standard Model degrees of freedom and can be found in any QFT textbooks, see for example \cite{QFTWeinberg,Pokorski}.
The potential reads as
\begin{eqnarray}
\label{Vphi}
V(H,\phi) = -m^2_1 H^{\dagger}H - m_2^2\phi^{\star}\phi + \lambda_1(H^{\dagger}H)^2 \nonumber \\
+ \lambda_2(\phi^{\star}\phi)^2  + 2\lambda_3(H^{\dagger}H)\phi^{\star}\phi. 
\end{eqnarray}
For the potential (\ref{Vphi}), if the vacuum expectation values are:
\begin{equation}
\begin{array}{lcr}
\label{vphi}
\sqrt{2} \langle H_i \rangle = v_H \delta_{i2},& &  \sqrt{2} \langle\phi\rangle = v_{\phi},
\end{array}
\end{equation}
then the tree-level mass parameters are:
\begin{eqnarray}
\label{mHiggs}
m_1^2 = \lambda_1 v_H^2 + \lambda_3 v_{\phi}^2, \\
\label{mphi}
 m_2^2 = \lambda_3 v_H^2 + \lambda_2 v_{\phi}^2.
\end{eqnarray}
Hence the model possesses two particles of masses: $m_1, m_2$, where $m_1$ is identified with the Higgs particle mass. The potential is bounded from below if:
\beq
\begin{array}{lcr}
\label{Vbounded}
\lambda_1> 0, & \lambda_2 > 0 & \lambda_3 > - \sqrt{\lambda_2\lambda_1}.
\end{array}
\eeq
If, in addition to (\ref{Vbounded}) the $\lambda_3 < \sqrt{\lambda_1\lambda_2}$ holds, then (\ref{vphi}) is the global minimum of $V$. 
For given $\lambda_1,\lambda_2, \lambda_3$ one can calculate $v_{\phi}$ and $m_2$ at tree level using (\ref{mHiggs}) and at loop level by changing the renormalisation schemes, because $v_H$ is known experimentally and the Higgs mass resulting from AS can be calculated, when we take $\lambda_3=0$. With $\lambda_0 = \frac{1}{2} \frac{m_1^2}{v_H^2} \approx 0.13$, one can parametrize the deviation from SM as: 
\beq
\label{tanbeta}
\tan \beta = \frac{\lambda_0-\lambda_1}{\lambda_3} \frac{v_{H}}{v_{\phi}}.
\eeq
We assume that its value is restricted by $|\tan \beta|<0.35$ what is the limit allowed by the present LHC data \cite{Lewandowski2017}. This constraint will be used to narrow down the possible values of $m_2$. 
In the Conformal Standard Model the coupling constants $Y^M_{ij}$ and $Y^{\nu}_{ij}$ are introduced, which are responsible for interactions of right-chiral neutrinos:
\beq
\mathcal{L}_Y= \frac{1}{2}Y^M_{ji}\phi N^{j\alpha}N_{\alpha}^i + Y^{\nu}_{ji}N^{j\alpha}H^{\top}\epsilon L_{\alpha}^i + \mathcal{L}^{SM}_Y +\textrm{h.c.},
\eeq 
where $\mathcal{L}_Y^{SM}$ is Yukawa part of the Standard Model Lagrangian part and $\epsilon$ is the antisymmetric $SU(2)_L$ metric. Following \cite{Lewandowski2017} we assume the degeneracy of Yukawa couplings $Y^M_{ij} = y_M \delta_{ij}$, which amplifies the CP violation and makes the resonant leptogenesis scenario possible, see \cite{Latosinski2015,Lewandowski2017,PILAFTSIS2004303} for details. The masses of right-chiral neutrinos are given by: 
\beq
M_N = y_M v_{\phi}/\sqrt{2},
\eeq
for leptogenesis to take place, one requires: $M_N > m_2$, so that the heavy neutrinos can decay. Moreover the Conformal Standard Model introduces a phase of the second scalar particle, called minoron, which can be a potential dark matter candidate, with mass $v^4/M_P^2$ originating from quantum gravity effects, where $v$ is some new parameter with $v \sim v_{\phi}$. The CSM beta functions  in the $\overline{MS}$-scheme, where $\hat{\beta}_{CSM} = 16\pi^2\beta_{CSM}$, are given by \cite{Lewandowski2017}:
\beq
\begin{array}{ll}
\hat{\beta}_{g_1}& =  \frac{41}{6} g_1^3, \textrm{   }  \hat{\beta}_{g_2} = - \frac{19}{6} g_2^3, \textrm{   }  \hat{\beta}_{g_3} = - 7 g_3^3,\\
\hat{\beta}_{y_t} &=  y_t\left(\frac{9}{2} y_t^2 - 8 g_3^2 - \frac{9}{4} g_2^2 - \frac{17}{12} g_1^2\right), \\
\hat{\beta}_{\lambda_1} &=  24 \lambda_1^2 + 4 \lambda_3^2 - 3 \lambda_1\left( 3 g_2^2 + g_1^2 - 4y_t^2\right)\\ 
&+\frac{9}{8} g_2^4 + \frac{3}{4} g_2^2 g_1^2 + \frac{3}{8} g_1^4 - 6 y_t^4,\\
\hat{\beta}_{\lambda_2} &= \left(20 \lambda_2^2 + 8 \lambda_3^2 + 6 \lambda_2 y_M^2 - 3y_M^4\right),\\
\hat{\beta}_{\lambda_3} &= \frac{1}{2}\lambda_3\left[ 24 \lambda_1 + 16 \lambda_2 +16 \lambda_3 \right.\\
&\left.- \left(9g_2^2 +3g_1^2\right) + 6 y_M^2 + 12y_t^2\right], \\
\hat{\beta}_{y_{M}} &=  \frac{5}{2} y_M^3, 
\end{array}
\eeq
where $g_1, g_2, g_3$ are  $U(1), SU(2), SU(3)$  Standard Model gauge couplings respectively, $y_t$ is the top Yukawa coupling. Following \cite{Lewandowski2017} we don't take into account the running of $Y^{\nu}$. If one takes $y_M = 0$ then after the redefinitions of the couplings the CSM beta functions reduce to the Higgs portal ones with  \cite{Higgsportal,Branco:2011iw,Gunion:2002zf,Wells:2009kq,Gong:2012ri,Lebedev:2011aq}. 
\subsection{Asymptotic safety and gravitational corrections}
The Weinbergs' notion of asymptotic safety (AS) \cite{Weinberg} can be summarised by his quote: ``A theory is said to be asymptotically safe if the essential coupling parameters approach a fixed point as the momentum scale of their renormalisation point goes to infinity.'' Let us assume that $\textbf{g}$ is a set of all the couplings of a theory and let $\textbf{g}|_{\mu_0}$ be a given a set of initial conditions at some momentum scale $\mu_0$. In our case we take: $\mu_0 =173.34$ GeV, see \cite{Laulumaa:2016ruk,Buttazzo}. These initial conditions together with the set of equations for running of couplings 
\beq
\beta_i(\textbf{g}(\mu)) = \mu \frac{\partial}{\partial \mu} g_i(\mu),
\eeq
describe completely and uniquely the behaviour of a physical theory. If for some $g_i$ we have $\beta_i(\textbf{g}^{\ast})=0$, then we call this $\textbf{g}^{\ast}$ a fixed point of  the $i$-th equation. The stable fixed points are called attractors, the unstable one are called repellers. If $\lim\limits_{\mu \to +\infty} \textbf{g}^{\ast} \equiv 0$ for all the couplings $g_i$ we call such a point a Gaussian fixed point. Theories where the couplings posses a Gaussian fixed at UV scales are called asymptotically free. Otherwise, when $g^{\ast}\neq 0$ we call such fixed point non-Gaussian/interacting and such theories are called asymptotically safe. If equation for running of the coupling $g_i$ possesses an unstable fixed points at UV scale, then there is only one low energy initial initial condition $g_i|_{\mu_0}$ per repeller such that the theory is fundamental up to the UV scale. \\ 
On the other hand gravity cannot be perturbatively quantized. Nevertheless one can utilize the effective field theory approach for energies below Planck scale to determine predictions of quantum gravity.
In particular the Standard Model (and its extensions) $\beta$ functions are modified with the gravitational corrections at high energies \cite{ShaposWetterlich,Robinson:2005fj,Zanusso:2009bs,Griguolo:1995db,Percacci:2003jz,Eichhorn:2017als,Eichhorn:2017egq,Laulumaa:2016ruk}:
\beq
\beta_i(\textbf{g}) = \beta_i^{\textrm{matter}}(\textbf{g}) + \beta_i^{\textrm{grav}}(\textbf{g},\mu). 
\eeq
The gravitational contributions to the beta functions acquire the general form for all the matter couplings \cite{ShaposWetterlich,Robinson:2005fj,Eichhorn:2017als,Eichhorn:2017egq,Laulumaa:2016ruk}:
\beq
\label{betagrav}
\beta^{\textrm{grav}}_i(\textbf{g},\mu) = \frac{a_i}{8\pi}\frac{\mu^2}{M_P^2+ \xi_0 \mu^2} g_i,
\eeq
due to the universal character of gravitational interactions. The $M_P = (8\pi G_N)^{-1/2} = 2.4 \times 10^{18}$ GeV is the low energy Planck mass.  The $\xi_0$ is some dimensionless constant.  Based on the results Functional Renormalisation Group investigation of pure gravity \cite{ShaposWetterlich,Reuter:1996cp,Percacci:2003jz,PhysRevD.68.044018} its value is taken as $\xi_0 = 0.024$. The $a_i$ are dimensionless constants and can be calculated for a given coupling $g_i$. 
According to \cite{Laulumaa:2016ruk,Zanusso:2009bs,Robinson:2005fj} one have $a_{g_1}=a_{g_2}=a_{g_3} =-1$ and $a_{y_t} = -0.5$ at the one-loop level. Since $y_M$ is a gauge coupling we assume that also $a_{y_M}=-1$.  The value $a_{\lambda_1}=+3$ is based on \cite{ShaposWetterlich,Laulumaa:2016ruk,PhysRevD.68.044018,Narain:2009fy}. For simplicity we assume that all of the $a_{\lambda_i}$ have the same absolute value, which can be supported by the calculations of the $a_{\lambda}$ parameters done for the Higgs Portal Models \cite{Einhorn}. However this calculation doesn't take into account fermions or higher order operators for gravity. For example the theory with the $R^2$ term in the gravity lagrangian can be identified with the scalar tensor theory of gravity \cite{Bezrukov,Starobinsky:1979ty} which is used to describe inflation and this term has negative $a_{R^2}$ \cite{Eichhorn:2018yfc} and so does the scalar. So we investigate all the possibilities in our article. To show that the absolute value of $a_{\lambda_2}, a_{\lambda_3}$ isn't an important factor when concerning the possible masses we have scanned over several values lying in the interval $|a_{\lambda_2}|, |a_{\lambda_3}| \in [1,3] $. Its change affects the allowed mass very weakly, of the order of $\pm 2$ GeV, this is because the gravitational corrections are heavily suppressed below the Planck scale. Indeed, the sign of $a_i$ is much more important than its exact value because the positive sign corresponds to the unstable Gaussian fixed point while the negative to the stable one. In result we investigate the four possibilities: $a_{\lambda_2},a_{\lambda_3} = \pm 3$. \\
The asymptotic safety assumption for quantum gravity allows us treat (extensions of) the Standard Model as fundamental (UV complete) only under the condition that the running of coupling constants doesn't possess any pathological behaviour up to the Planck scale. It imposes two conditions \cite{ShaposWetterlich,Lewandowski2017}. We will call them asymptotic safety (AS) conditions. Firstly, there should be no Landau poles up to the Planck scale. Secondly, the electroweak vacuum should be stable for all scales:
\beq
\begin{array}{lcr}
\label{Vacuum}
\lambda_1(\mu) > 0, & \lambda_2(\mu) >0, & \lambda_3(\mu) > - \sqrt{\lambda_2(\mu)\lambda_1(\mu)}.
\end{array}
\eeq
 The second condition comes from the assumption that there is essentially no new physics between EW scale and Planck scale, despite the one described by Conformal Standard Model. Obviously at Planck scale all the matter couplings goes to zero, and hence all the beta functions goes to zero. The next paragraph is dedicated to the calculation of the lambda-couplings ($\lambda_1, \lambda_2, \lambda_3, y_M$) satisfying these conditions. 
\section{Calculation of lambda couplings}
In this paragraph we calculate the set of allowed lambda-couplings satisfying the asymptotic safety conditions. If not specified otherwise, the value of a coupling (for example on the plots below) means its value at $\mu_0 = 173.34$. In the low energy regime the graviton loops can be neglected \cite{ShaposWetterlich,SBCS, SBCS2} and they become important near the Planck scale and manifest in the form of the gravitational corrections to the $\beta$ functions. The gauge and Yukawa couplings renormalisation group equations dynamics is not affected by the running of $\lambda_i$ and $y_M$. The low-energy values at $\mu_0 = 173.34$ are taken as \cite{Laulumaa:2016ruk}: $g_1(\mu_0)=0.35940$, $g_2(\mu_0)= 0.64754$, $g_3(\mu_0)=1.1888$ and $y_t(\mu_0)= 0.95113$. Then the evolution of these couplings with energy is obtained. The running of $y_M$ is also independent from all other couplings. So far the experimental value for $y_M$ is unknown, so to obtain the allowed $\lambda_i (y_M)$ one has to scan over all possible values of $y_M$. Using the AS conditions we have calculated the allowed interval of $y_M$ for $a_{y_M} =-1$ as $y_M \in [ 0.0, 0.925]$, for $a_{y_M}=+1$ we have $y_M=0.0$. We plug the evolution of the gauge and Yukawa couplings and each allowed $y_M$ into coupled equations for $\lambda_1, \lambda_2, \lambda_3$. Moreover, the magnitudes of $a_i$ coefficients are such that the theory becomes asymptotically free near the Planck scale, see \cite{ShaposWetterlich} for further details, which justifies the use of perturbation approach. Furthermore we expect that if the cosmological constant runs then it does not affect the matter couplings below the Planck scale. This reasoning is supported by the fact that both Higgs mass \cite{ShaposWetterlich} and mass difference between top and bottom quark were accurately predicted \cite{Eichhorn:2018whv} without taking into account the running of cosmological constant. 
Furthermore in the case of the unimodular gravity \cite{Eichhorn:2013xr}, which is equivalent to the Einstein theory at the classical level, the cosmological constant isn't a dynamical degree of freedom hence it doesn't run at all. These arguments suggest that we don't take the cosmological constant running into account. As a result we are looking for the sets of initial values $ y_M, \lambda_i$ such that they will all drop to zero near the Planck scale.
\subsection{Coefficients: $a_{\lambda_3} = -3, a_{\lambda_2} = -3$}
\begin{figure}[h]
\centering
\includegraphics[width=1.0\linewidth]{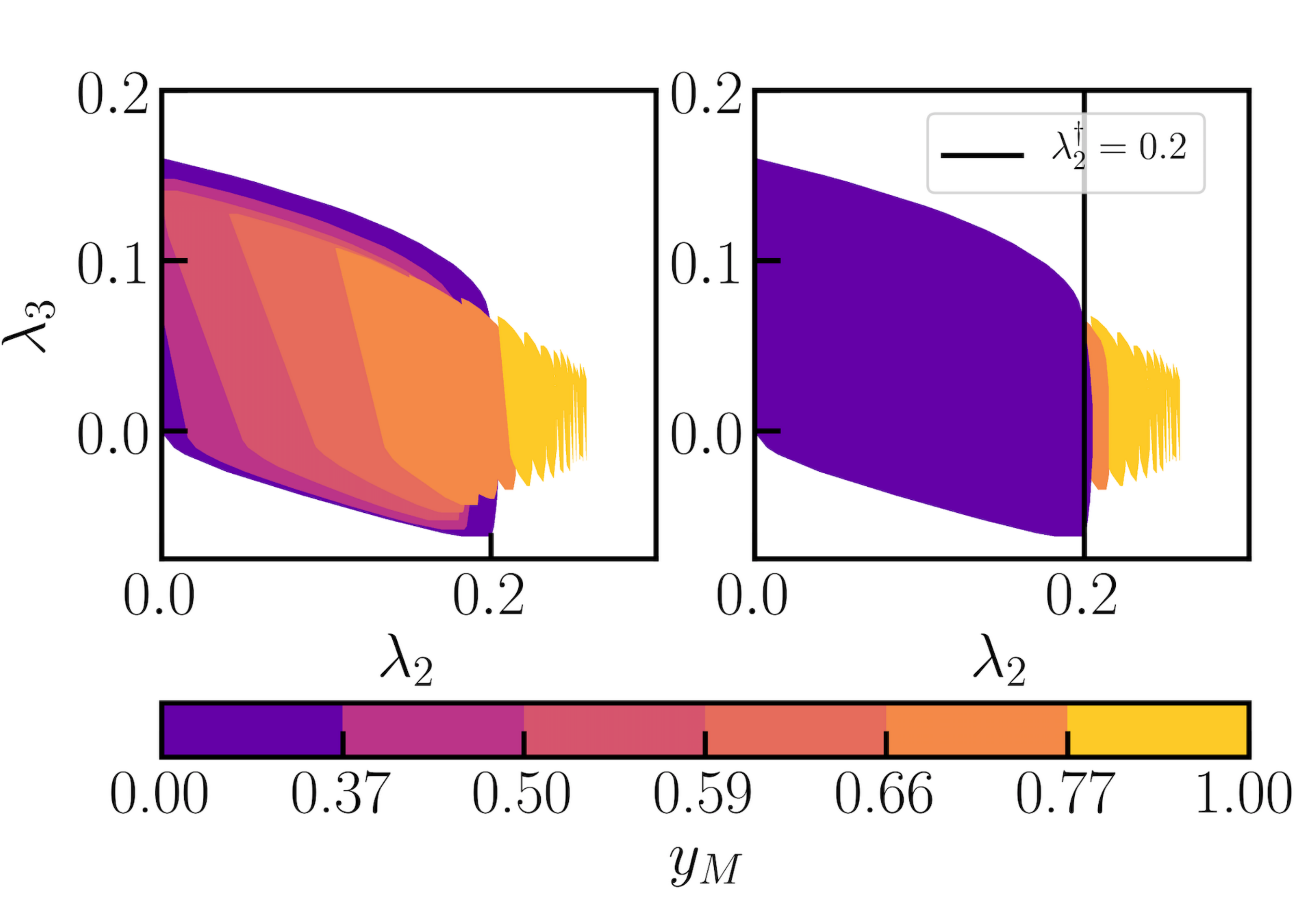}
\caption{Maximal (left) and minimal (right) $y_M(\lambda_3,\lambda_2)$, $a_{\lambda_2}=-3, a_{\lambda_3}=-3$}
\label{Fig1}
\end{figure}

\begin{figure}[!h]
\centering
\includegraphics[width=1.0\linewidth]{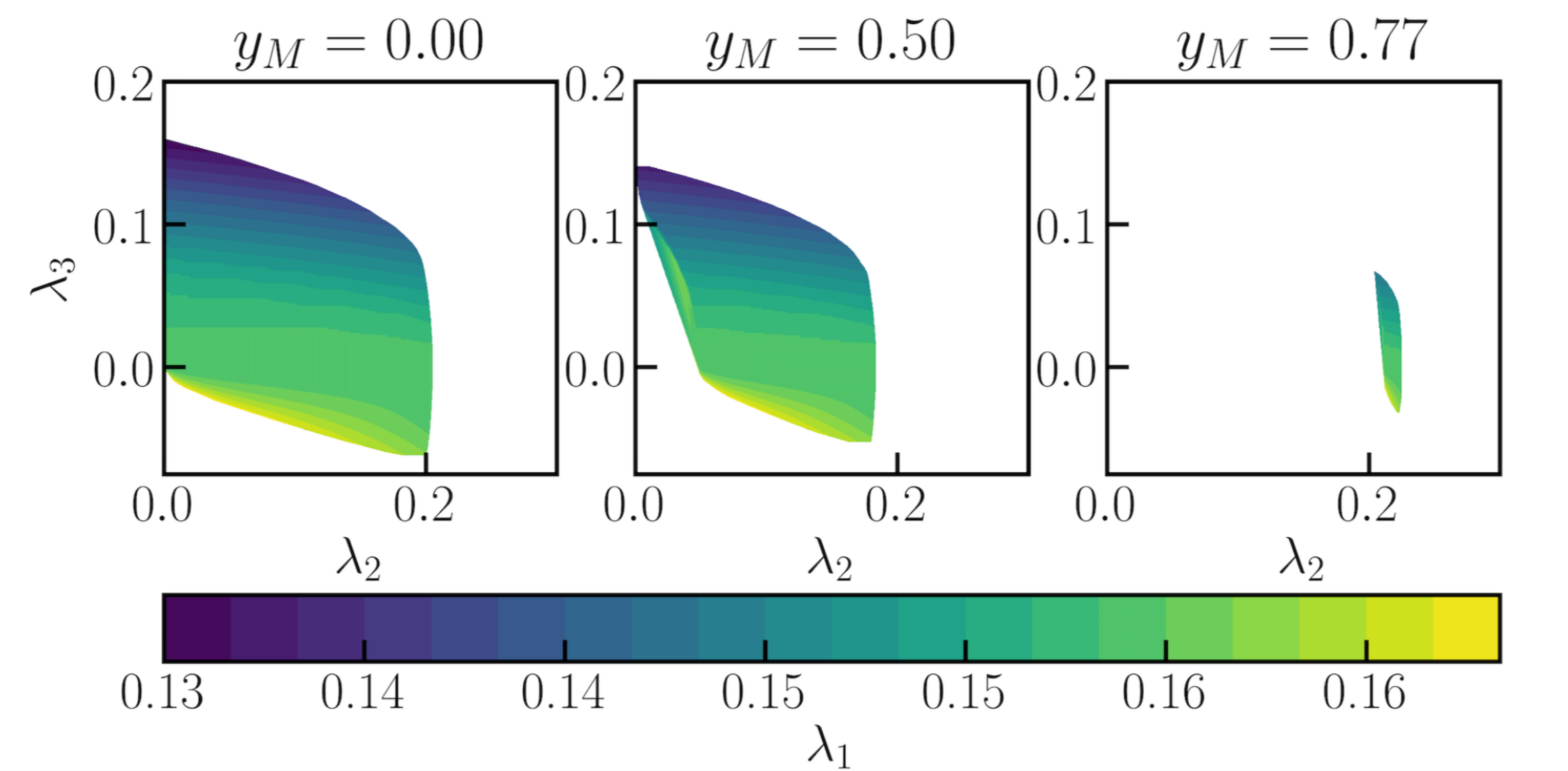}
\caption{Values of $\lambda_1$ for $y_M=0.0$  (left) \\ and $y_M=0.5$ (mid), $y_M=0.77$ (right)}
\label{lambda1plot}
\end{figure}

We start with the most general case, when $a_{\lambda_2} = a_{\lambda_3} = -3$. Since for each possible combination of $\lambda_2, \lambda_3, y_M$ only one $\lambda_1$ is allowed, then we consider a set $\mathfrak{M}$ of allowed $\lambda_2, \lambda_3, y_M$ ($\lambda_2$ and $\lambda_3$ depend mutually on each other and on $y_M$),  and there is a $\lambda_1 (\lambda_2, \lambda_3, y_M)$ assigned to each of the points of this set. This set looks roughly like a sea wave, where $y_M$ is the height. On the Fig.~\ref{Fig1} we show the surfaces of maximal and minimal possible $y_M$ all the points in between are also in the set $\mathfrak{M}$.
\FloatBarrier
\FloatBarrier
On the right figure there is a special line $\lambda_2^{\dagger} =0.2$. Namely, for the region where $\lambda_2 < \lambda_2^{\dagger}$ the minimal $y_M$ is $0.0$, while when $\lambda_2> \lambda_2^{\dagger}$ the minimal and maximal values are very close to each other: $y_M(\textrm{max})-y_M(\textrm{min}) \sim 0.01$. This behaviour can be explained by the observation that $\lambda_2> \lambda_2^{\dagger}$ requires $y_M>0.70$ to keep $\hat{\beta}_{\lambda_2}$ small enough throughout the evolution. As we have checked in all other cases the sets of allowed couplings  form a 1D or 2D subsets of $\mathfrak{M}$. \\
The $\lambda_1$ renormalisation group equation is affected directly only by $\lambda_3$, however the running of $\lambda_3$ depends on $\lambda_2$ and $y_M$. On the Fig.~\ref{lambda1plot}
we show the $\lambda_1$ dependence on other couplings for three chosen $y_M$ as an example.
\FloatBarrier
\FloatBarrier
\subsection{\label{casetwo}Coefficients: $a_{\lambda_3} = -3, a_{\lambda_2} = +3$}
\begin{figure}[!h]
\centering
\includegraphics[width=1.0\linewidth]{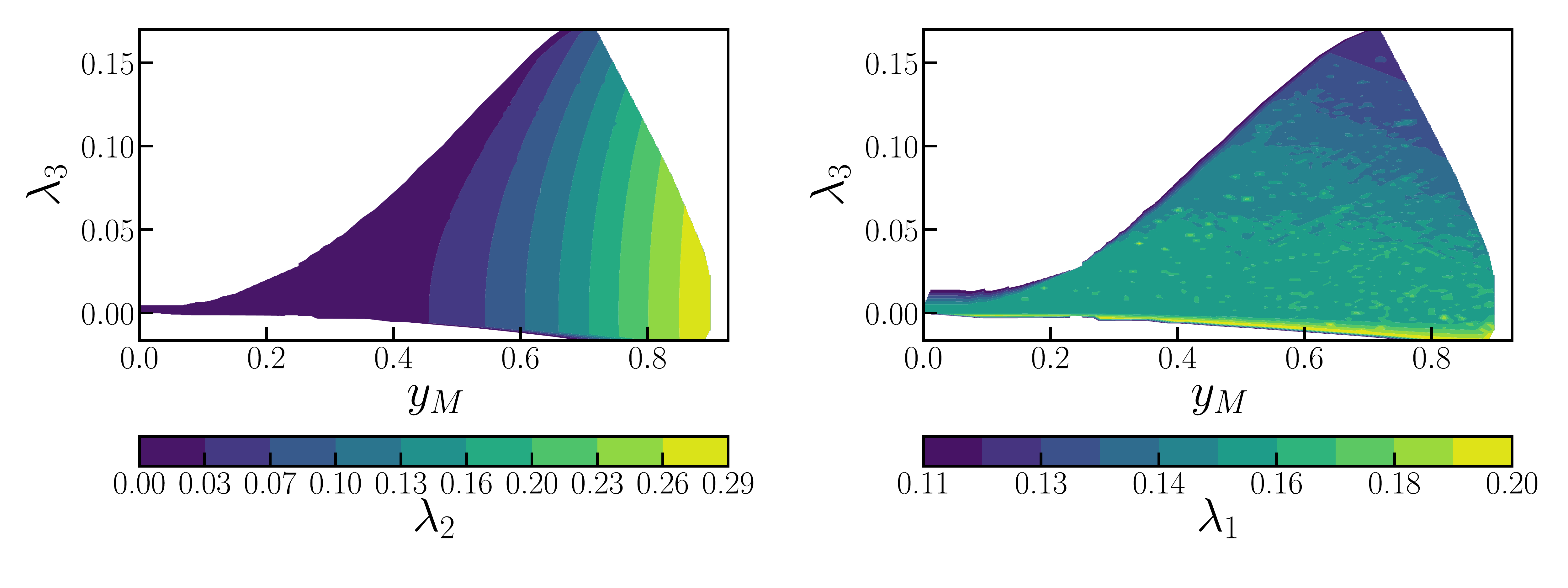}
\caption{$\lambda_1$ dependence on $\lambda_3$ and $y_M$, $a_{\lambda_2}=+3, a_{\lambda_3}=-3$}
\label{Fig3}
\end{figure}

In this case there is one allowed $\lambda_1$ and $\lambda_2$ per set of $\lambda_3$ and $y_M$ satisfying  the AS conditions.
On the Fig.~\ref{Fig3} we present this dependence. As we can see there is much less spread in possible values of $\lambda_1$ than $\lambda_2$. This is because all the gauge coupling initial values are fixed for $\lambda_1$, which is not the case for $\lambda_2$. 
\FloatBarrier
\FloatBarrier
For this case the domain of allowed $\lambda_3$ and $y_M$ is smaller than for $\mathfrak{M}$. However for this domain we have:  
$\lambda_2(a_{\lambda_3}=+3, \lambda_3, y_M) = \min_{\lambda_2} \left(\lambda_2(a_{\lambda_2}=-3,\lambda_3, y_M)\right\}.
$
\subsection{Coefficients: $a_{\lambda_3} = +3$, $a_{\lambda_2} = \pm 3$}
In this case, we found that only $\lambda_3(\mu_0)=0$ satisfies the AS conditions, hence $\lambda_3$ is zero at all energy scales. Then the $\phi$ sector is decoupled from the rest of Standard Model, which makes it a scalar dark matter candidate, like in \cite{Einhorn}. The numerical solution for $\lambda_1$ gives $\lambda_1=0.1537$, which agrees the Standard Model predictions from asymptotic safety, see \cite{ShaposWetterlich,Laulumaa:2016ruk}. The allowed region for $\lambda_2$ is determined by the AS conditions for this coupling and depends heavily on $y_M$.  
\FloatBarrier
\begin{figure}[!h]
\centering
\includegraphics[width=1.0\linewidth]{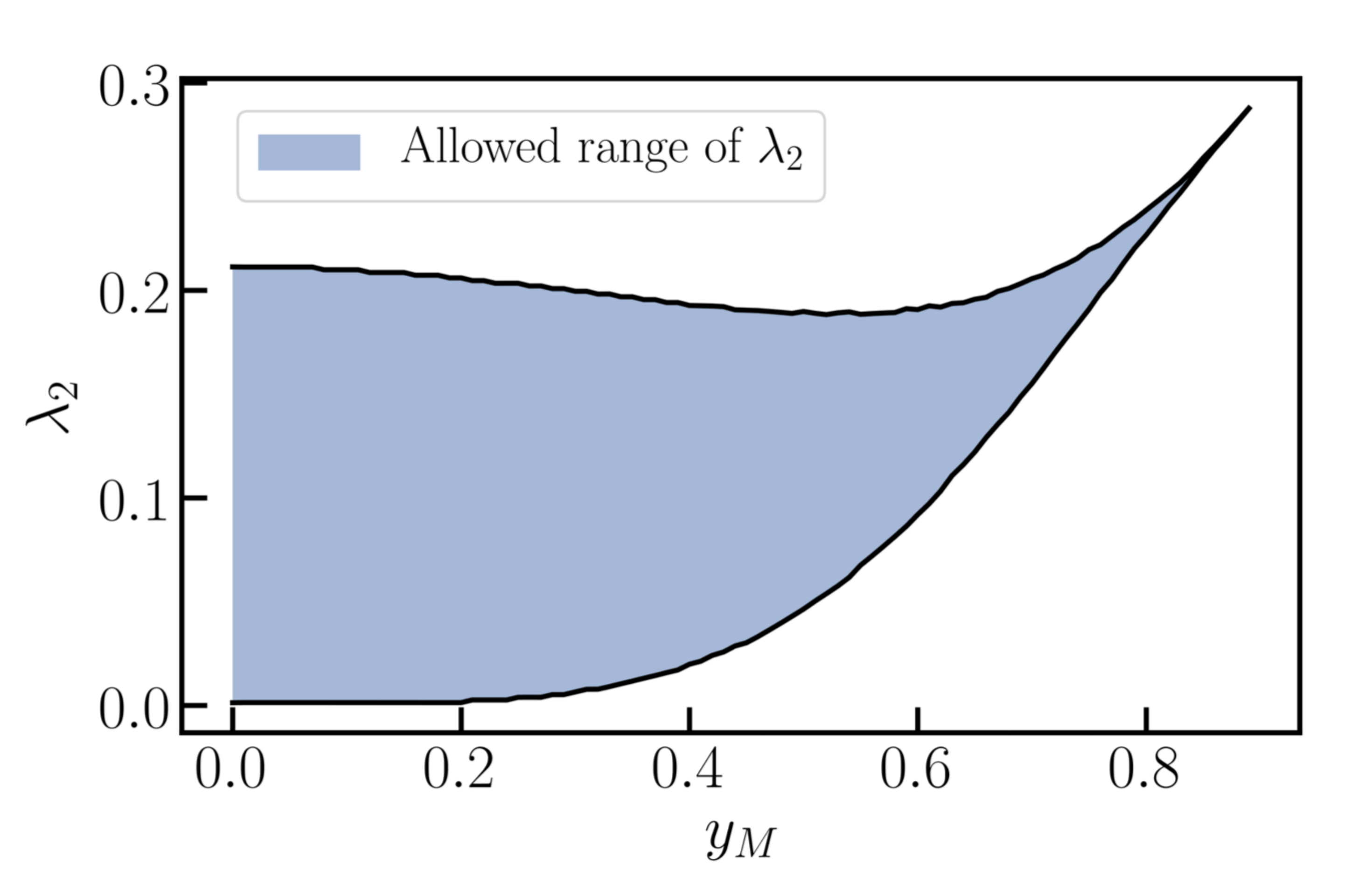}
\caption{Allowed range of $\lambda_2$ coupling, $a_{\lambda_2}=-3, a_{\lambda_3}=+3$}
\label{Figcasethree}
\end{figure}
\FloatBarrier
We checked that the allowed value for $\lambda_2$ shown on the  Fig.~[\ref{Figcasethree}] is the subset of $\mathfrak{M}$ with the condition $\lambda_3=0.0$. We have also calculated that $\lambda_2(a_{\lambda_2} =+3,y_M)$ mimics the lower bound for $\lambda_2 (a_{\lambda_{2}} =-3,y_M)$. Moreover this bound is the same as for the running of $\lambda_2$ without gravitational corrections. 
\section{The second scalar mass}
In this paragraph we calculate the $v_{\phi}, m_2$ and $M_N$. For the decoupled (Standard Model) case one obtains $\lambda_1 = 0.1537$ hence the the Higgs mass is given by: 
\beq
m_1 = \sqrt{2\lambda_1 v_{H}^2} \approx 136 \textrm{GeV},
\eeq

where the $\lambda_1$ is calculated from the AS requirements and is in $\overline{MS}$ scheme.
So one can see that the effects regarding the renormalisation of mass and the ones concerning the changing of renormalisation scheme (from $\overline{MS}$ to the physical one) have the contribution of order of $10$ GeV to the masses calculated at the tree level relations (without renormalisation of mass), which was demonstrated in \cite{ShaposWetterlich,Laulumaa:2016ruk}. Hence due to other much bigger sources of uncertainties, like the value of $y_M$ we assume that the CSM tree level relations $(\ref{mHiggs}, \ref{mphi})$ hold. This assumption restricts $\lambda_3$ to be nonnegative, otherwise one would obtain negative values for $v_{\phi}^2$. sufficient for the calculations of the allowed $\lambda$ couplings.  By comparing the one loop \cite{Laulumaa:2016ruk} and two-loop calculation \cite{ShaposWetterlich} of $\lambda_1$, where the outcome is almost identical, we conclude that higher than the first non-trivial order is  sufficient for our purpose. In order to calculate $v_{\phi}$ and $m_2$ at tree level we take $v_{H}=246$ GeV, then we solve the equations (\ref{mHiggs}, \ref{mphi})
with $m_1=136$ GeV. Obviously we are able to predict the $m_2$ and $v_{\phi}$ only in the case when the second scalar sector is coupled to the SM sector. Hence, we restrict only to the case when: $\lambda_3 \neq 0$. For small $\lambda_3$ tiny changes in $\lambda_1$ results in enormous changes in $v_{\phi}$, so we treat $\lambda_3 <0.01$ as a decoupled case. Our claim that $\lambda_3$ should be large enough can also be justified with the LHC condition $ |\tan \beta| < 0.35$, because small $\lambda_3$ results in large $\beta$. In our analysis we have excluded all the sets of parameters not satisfying the LHC and the global stability conditions (at $\mu_0$: $\sqrt{\lambda_1(\mu_0) \lambda_2(\mu_0)} > \lambda_3(\mu_0)$). After doing that we have two separate cases: $a_{\lambda_2 }=+3$ and $a_{\lambda_2 }=-3$.
Moreover if the second scalar particle is unstable (following CSM \cite{Lewandowski2017}), then at the tree level we have: $2m_1 < m_2$. Also the $y_M=0.0$ situation is an interesting situation, when CSM reduces to the Higgs portal case. The masses of right handed neutrinos are can also be calculated, $M_N = y_M v_{\phi}/\sqrt{2}$ in case when $y_M\neq 0$. Below we illustrate these combinations of possible conditions.
\FloatBarrier
\begin{table}[h!]
\caption{Arrays of allowed masses}
\begin{tabular}{|c|c|c|c|c|c|}
\hline
$a_{\lambda_2}$ & $2m_1<m_2 $ & $y_M$ &  $m_2$ [GeV]& $v_{\phi}$ [GeV]& $M_N$ [GeV]  \\
\hline
$+3$ & yes & $0.84$ & $275 $ & $538$ & $319  $\\
\hline
$+3$ & yes & $0.85$ & $296 $ & $574 $ & $345 $ \\
\hline
\hline
$-3$ & yes &  $0.77_{-0.06}^{+0.07}$ & $300^{+28}_{-28} $ & $586^{+60}_{-46} $ & $342^{+41}_{-41} $ \\
\hline
$-3$ &  no &  $0.00$ & $160^{+103}_{-100} $ &  $300^{+275}_{-15} $ & NA \\
\hline
\end{tabular}
\end{table}
\FloatBarrier
In the second row ``yes'' means that we've taken into account this condition, while ``no'' means the opposite. As we can see for $y_M=0.0$ the second scalar mass is $m_2$ is smaller than $ 2 m_1$ making this particle stable. In the Conformal Standard Model case the right-chiral neutrinos turns out to be unstable ($ m_2 < y_M v_{\phi}/\sqrt{2}$ holds for each of the sets of parameters) making the leptogenesis scenario possible. Furthermore if we assume that $v = v_{\phi}$, then the mass of the minoron is given by: $v_{\phi}^2/M_P \approx 10^{-4}$ eV, which is of the same order as estimated in \cite{Lewandowski2017}.  

We have also analyzed the running of the $\beta$ functions for $m_2$ and $m_1$, where we took $a_{m_i}= -1$. It gives no new bounds on $m_2$ and lambda-couplings. This result is in close relation with scenario B analysed in \cite{Einhorn}, however in our case the $\phi$ particle isn't decoupled from SM. We have also taken into account the restrictions from  quadratic divergences cancellation, which can be achieved if Softly Broken Conformal Symmetry \cite{SBCS} requirements are satisfied (these requirements in CSM are not essentially changed even after adding gravitational corrections \cite {SBCS2}). Apparently it gives no new restrictions for the parameters additional to the asymptotic safety conditions. Moreover the SBCS requirements are satisfied at $E=M_P$ with all the couplings becoming asymptotically free, which was suggested in \cite{SBCS}.
\section{Conclusions}
Our investigation shows that the Conformal Standard Model has a set of parameters compatible with the asymptotic safety conditions, so supplemented with gravitational corrections it can be a UV fundamental theory and \cite{Lewandowski2017}: ``allows for a comprehensive treatment of all outstanding problems of particle physics.'' Moreover, CSM can be slightly modified to incorporate an inflation scenario \cite{Lebedev:2011aq,Kwapisz}, which agrees with 2013 Planck data analysis \cite{Ade:2015tva,Guth:2013sya,Kaiser:2015usz}. Inflation can also take place due to the asymptotically safe scenario \cite{Weinberg:2009wa}.

The restrictions on the couplings and masses derived in this article allow us to make theoretical predictions for free parameters of the models extending the Standard Model. The lambda-couplings can be directly measured (or explicitly calculated from masses of the new scalar particles) in LHC or indirectly measured in cosmological observations. In particular our investigation supports the claim \cite{Meissner:2012hd} that the excess of events with four charged leptons at $E \sim 325$ GeV seen by the CDF \cite{Aaltonen:2011jn} and CMS \cite{Chatrchyan:2012dg} Collaborations can be identified with a detection of a new `sterile' scalar particle proposed by the Conformal Standard Model. On the other hand the compatibility of the LHC run 1 data with the heavy scalar hypothesis was investigated in \cite{LHCrun1,LHCpheno}. The hypothetical heavy boson mass is measured to be around $272$ GeV (in the $270-320$ GeV range). Hence we would like to emphasise that this experimental analysis agrees with the theoretical range provided by our calculations and with the claim discussed in \cite{Meissner:2012hd} at least up to the order of several GeV. The fact that $m_2>2m_1$ implies the $y_M \neq 0.0$ underlies the role of right-handed neutrinos not only in context of smallness of neutrino masses, but also in the case of asymptotically safe BSM physics, which makes the Conformal Standard Model unique among the 2HDM and Higgs portal models.
 
We hope that our analysis will be helpful in search for new particles at LHC and future colliders and the new scalar particle predicted by this model can be detected in the nearby future. 
\begin{acknowledgments}
We thank Piotr Chankowski for valuable discussions. J.H.K. would like to thank Yukawa Institute for Theoretical Physics for hospitality and support during this work. The PL-Grid Infrastructure is gratefully acknowledged. K.A.M. was 
 partially supported by the Polish National Science Center grant DEC-2017/25/B/ST2/00165. J.H.K. was supported by the National Science Centre, Poland grant 2018/29/N/ST2/01743.
\end{acknowledgments}  
\addcontentsline{toc}{section}{The Bibliography}
\bibliography{mybibfile.bib}{}
\bibliographystyle{apsrev4-1}
\end{document}